\documentclass[prl,aps,twocolumn,showpacs,floatfix]{revtex4}
\usepackage{graphicx}
\usepackage{bm}

\begin{document}

\title{Optical Flux Lattices for Ultracold Atomic Gases}

\author{N. R. Cooper}
\affiliation{Cavendish Laboratory, University of Cambridge, J.~J.~Thomson Ave., Cambridge CB3~0HE, U.K.}

\begin{abstract}

  We show that simple laser configurations can give rise to ``optical
  flux lattices'', in which optically dressed atoms experience a
  periodic effective magnetic flux with high mean density.  These
  potentials lead to narrow energy bands with non-zero Chern
  number. Optical flux lattices will greatly facilitate the
  achievement of the quantum Hall regime for ultracold atomic gases.

\end{abstract}
\date{January 12, 2011}
\pacs{ 67.85.Hj, 67.85.-d, 37.10.Vz}


\maketitle

One of the most important techniques in the ultracold atom toolbox is
the optical lattice\cite{blochdz}: a periodic scalar potential formed
from standing waves of light.  Optical lattices are central to the use
of atomic gases as quantum simulators, and allow the exploration of
strong-correlation phenomena related to condensed matter
systems\cite{jakschbh,greiner}.  Their usefulness derives from the
fact that the scalar potential has a lengthscale, set by the optical
wavelength $\lambda$, that is similar to the typical interatomic
spacing.

Largely separate have been the experimental developments of effective
vector potentials, representing the coupling of a charged particle to
a magnetic field.  A uniform magnetic field can be simulated for
neutral atoms using rotation\cite{blochdz,fetter,advances}. Gauge
fields leading to an effective magnetic field can also be generated by
optical dressing\cite{dalibardreview}.  These techniques have allowed
experimental studies of quantized vortices in condensed Bose and
(paired) Fermi
gases\cite{MadisonCWD00,Zwierlein,spielmanfield,blochdz,fetter}.  An
interesting regime of strong correlation, with connections to the
fractional quantum Hall effect, is expected when the effective
magnetic flux density $n_\phi$ is sufficiently high that the
lengthscale ${n_\phi}^{-1/2}$ is comparable to the interatomic
spacing\cite{cwg,advances}.  However, the magnetic flux densities
achievable using existing techniques are relatively small for large
systems\cite{SchweikhardCEMC92,spielmanfield,dalibardreview}, so this
strongly correlated regime occurs at very low densities when the gas
is weakly interacting and highly susceptible to uncontrolled
perturbations\cite{SchweikhardCEMC92,advances}.

In this paper we describe simple laser configurations that use optical
dressing to generate what we refer to as ``optical flux
lattices''. While a conventional optical lattice imprints a periodic
scalar potential, an optical flux lattice also imprints {\it a
  periodic magnetic flux density with non-zero mean}, and large
magnitude $\bar{n}_\phi \sim 1/\lambda^2$.  We emphasize that the
magnetic flux density is a continuous function of position, so these
potentials are distinct from proposed ways to imprint gauge potentials
onto deep optical lattices which apply only in the tight-binding
limit\cite{jakschzoller,mueller,gerbier}; furthermore, optical flux
lattices require only a small number of lasers, so are much easier to
implement than these tight-binding proposals.  We show that optical
flux lattices lead to narrow bands with non-zero Chern
number\cite{thoulesschern}.  In particular the lowest energy band is
topologically equivalent to the lowest Landau level.  Since the
lengthscale $\bar{n}_\phi^{-1/2}\sim \lambda$ is similar to the
typical interatomic spacing, optical flux lattices will allow the
study of quantum Hall physics at high densities where interaction
energy scales are large.

We consider an atom moving in optical fields within the rotating wave
approximation, with Hamiltonian
\begin{equation} \hat{H} = \frac{\hat{{\bm
      p}}^2}{2m} \hat{I} + V \hat{M}({\bm r}) \label{eq:ham}
\end{equation}
where $V$ is the energy scale of the optical potential, of
dimensionless form $\hat{M}({\bm r})$.  We focus on two-level systems
and write
\begin{equation}
\hat{M} = \vec{M}({\bm r})\cdot \hat{\vec{\sigma}} = 
\left(\begin{array}{cc}
M_z & M_x - iM_y \\
M_x + iM_y & -M_z
\end{array}\right)
\end{equation}
with $\hat{\vec{\sigma}}$ the vector of Pauli matrices.  Additional
scalar potentials can be added by conventional optical lattices; for
simplicity, we neglect these here.  The off-diagonal terms ($M_{x,y}$)
arise from the Raman coupling that effects the interspecies
conversion. The diagonal term ($M_z$) represents a species-dependent
potential.  One possible implementation of the two-level system is
with the groundstate and long-lived excited state of an alkaline earth
atom or ytterbium\cite{gerbier}, in which case $M_z$ can be generated
by a laser at an ``anti-magic'' wavelength, $\lambda_{\rm am}$.  As we
shall describe below, optical flux lattices can be formed by a
standing wave at $\lambda_{\rm am}$ and three travelling waves of the
Raman laser\cite{Note1}.  We shall restrict attention to quasi
two-dimensional (2D) systems with ${\bm r} = (x,y)$.  The resulting
flux lattices are readily adapted to 3D, with net flux along one
direction.

The emergence of an effective gauge potential is best
understood when the kinetic energy is small compared to the energy
spacing of the (local) dressed states, obtained from the eigenvalues
of $V \hat{M}$. The atom then moves through space adiabatically,
staying in a given dressed state.  The adiabatic limit is always valid
for sufficiently large $V$ provided the spectrum of $\hat{M}$ is
non-degenerate. Assuming  this to be true (as shall be verified 
below for the cases of interest), we consider the adiabatic motion in
a normalized dressed state
\begin{equation}
|\Phi({\bm r})\rangle = 
\left(\begin{array}{c}
\phi_1({\bm r})\\
\phi_2({\bm r})
\end{array}\right)\,.
\label{eq:dressed}
\end{equation}
Projecting (\ref{eq:ham}) onto adiabatic motion on the state
(\ref{eq:dressed}) leads to an effective Hamiltonian with both a
scalar and a vector potential, the latter given
by\cite{dalibardreview}
\begin{equation}
q{\bm A} = i \hbar\langle \Phi | {\bm \nabla}\Phi\rangle
\label{eq:vecpot}
\end{equation}
for effective charge $q$.  The number density of magnetic flux quanta
perpendicular to the $xy$-plane is therefore
\begin{equation} n_\phi \equiv \frac{qB}{h} = \frac{q}{h}{\bm \nabla}\times{\bm
  A}\,.
\label{eq:flux} 
\end{equation}

For optical fields of wavelength $\lambda$, it is natural to assume
that the vector potential (\ref{eq:vecpot}) is smoothly varying with
$|q{\bm A}| \lesssim h/\lambda$. Then, the maximum mean flux density
in a region of space of sides $L_x,L_y\gg \lambda$ may be found by
applying Stokes' theorem: $ \int n_\phi d^2{\bm r} \equiv \bar{n}_\phi
L_xL_y = (q/h) \oint {\bm A}\cdot d{\bm r} \lesssim
(L_x+L_y)/\lambda$, leading to $\bar{n}_\phi \lesssim 1/L\lambda $
with $L=\mbox{min}(L_x,L_y)$.  All existing proposals for optically
induced gauge fields in the continuum\cite{dalibardreview}, and the
scheme implemented in Ref.\onlinecite{spielmanfield}, are of this form
with the scale $L$ set by the width of the cloud.  Since, typically,
$L\gg \lambda$, this leads to relatively small flux density,
$\bar{n}_\phi\sim 1/L\lambda$.

Although apparently very general, these considerations neglect the
fact that smoothly varying optical fields can induce singularities in
$q{\bm A}$. These singularities depend on the gauge used for
(\ref{eq:dressed}), and cause no singularities in gauge-invariant
properties. Such issues arise whenever a U(1) gauge field has non-zero
flux through a closed manifold, notably leading to Dirac strings for a
magnetic monopole.  For the optical flux lattices we propose here,
there is a net flux through a unit cell which (due to the spatial
periodicity) has the topology of a torus.  We avoid technical
difficulties of the gauge-dependent singularities by defining the
local Bloch vector
\begin{equation}
\label{eq:bloch}
\vec{n}({\bm r}) = \langle \Phi (\bm {r})| \hat{\vec{\sigma}} | \Phi({\bm r})\rangle
\end{equation}
for which $\vec{n}\cdot\vec{n}=1$.  The flux density is then
\begin{equation} n_\phi = - \frac{1}{8\pi}\epsilon_{ijk}\epsilon_{\mu\nu}n_i\partial_\mu
n_j\partial_\nu n_k \,.
\label{eq:topo} 
\end{equation}
This is (minus) the ``topological density'' of the map from position
space, ${\bm r} = (x,y)$, to the surface of the Bloch sphere,
$\vec{n}$\cite{girvinleshouches}.  The number of flux quanta through a
region $A$ is $\int_A n_\phi \,d^2{\bm r} = {\Omega}/{4\pi}$ where
$\Omega$ is the solid angle that region $A$ maps to on the Bloch
sphere.  Thus, each time the Bloch vector wraps the surface of the
sphere corresponds to one magnetic flux quantum.
Optical flux lattices are spatially periodic configurations for which
the Bloch vector wraps the sphere an integer number, $N_\phi$, times
in each unit cell.  The lattice vectors ${\bm a}_1$ and ${\bm a}_2$
are both of order the optical wavelength $\lambda$, so the mean flux
density is of order $\bar{n}_\phi \sim N_\phi/{\lambda^2}$ which is
large.
We focus on two cases of high symmetry which achieve this
goal.

{\it Square Lattice:} Consider the optical coupling
\begin{equation}
\hat{M}_{\rm sq} = 
\cos(\kappa x) \hat{\sigma}_x
+ \cos(\kappa y) \hat{\sigma}_y
+ \sin(\kappa x)\sin(\kappa y)\hat{\sigma}_z
\label{eq:vsq}
\end{equation}
where $\kappa \equiv 2\pi/a$.  This has square symmetry with ${\bm
  a}_1=(a,0), {\bm a}_2=(0,a)$.  Achieving this high symmetry in
experiment may involve tilting the lasers out of the $xy$-plane to
tune the periods of the Raman and the species-dependent
fields\cite{Note2}.
The eigenvalues are
non-degenerate at all positions, so the dressed states admit an
adiabatic limit.
\begin{figure}
\includegraphics[width=0.98\columnwidth]{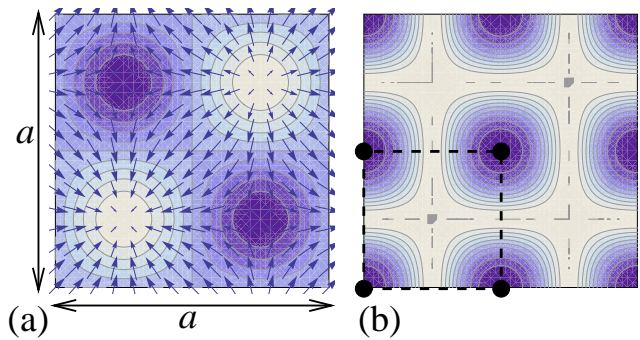}
\caption{ \label{fig:square} Properties of the lower energy dressed
  state for the square optical flux lattice (\ref{eq:vsq}), as a
  function of position in the unit cell. (a) Local Bloch vector, as
  represented by the vector $(n_x,n_y)$ and contours of $n_z$ (light
  shading for $n_z<1$ and dark for $n_z>1$).  (b) The local flux
  density $n_\phi$ is maximal on the lines $x= \pm a/4$ and $y=\pm
  a/4$ and vanishes at four points.  The lattice sites and
  nearest-neighbour hopping in the tight-binding limit are indicated
  by dark circles and dashed lines.  }
\end{figure}
Fig.\ref{fig:square}(a) shows that, for the lower energy dressed
state, the $(n_x,n_y)$ components of the Bloch vector exhibit two
vortices and two anti-vortices in the unit cell.  These vortices lead
to gauge-dependent singularities in the vector potential. However, the
Bloch vector varies smoothly, with $n_z = \pm 1$ at the vortex cores,
in such a way that all four of these regions wrap the sphere in the
same sense and contribute a flux density of the same sign. The flux
density is shown in Fig.\ref{fig:square}(c). It is everywhere
non-negative and has total flux $N_\phi=2$ in the unit cell. (This may
be seen by noting that the two vortices and two antivortices cause
$\vec{n}$ to wrap the Bloch sphere twice.)  The flux density is not
constant, and vanishes at four points in the unit cell.  These four
points coincide with the locations at which the adiabatic energy is
minimum, Fig.\ref{fig:square}(b), so form the lattice sites in the
tight-binding limit.

{\it Triangular lattice:} An optical flux lattice with triangular symmetry 
is generated by
\begin{equation}
\hat{M}_{\rm tri}
= \cos({\bm r}\cdot{\bm \kappa}_1) \hat{\sigma}_x
+ \cos({\bm r}\cdot{\bm \kappa}_2) \hat{\sigma}_y
+ \cos({\bm r}\cdot{\bm \kappa}_3) \hat{\sigma}_z
\label{eq:vtriang}
\end{equation}
where ${\bm \kappa}_1= (1,0)\kappa\, , {\bm \kappa}_2=
\left(\frac{1}{2},\frac{\sqrt{3}}{2}\right)\kappa$ and ${\bm
  \kappa}_3={\bm \kappa}_1-{\bm \kappa}_2$, with $\kappa \equiv
4\pi/(\sqrt{3}a)$, giving lattice vectors ${\bm a}_1 =
(\sqrt{3}/2,-1/2)a$ and ${\bm a}_2 = (0,1)a$.  Again, the eigenvalues
are non-degenerate at all positions.
\begin{figure}
\includegraphics[width=0.95\columnwidth]{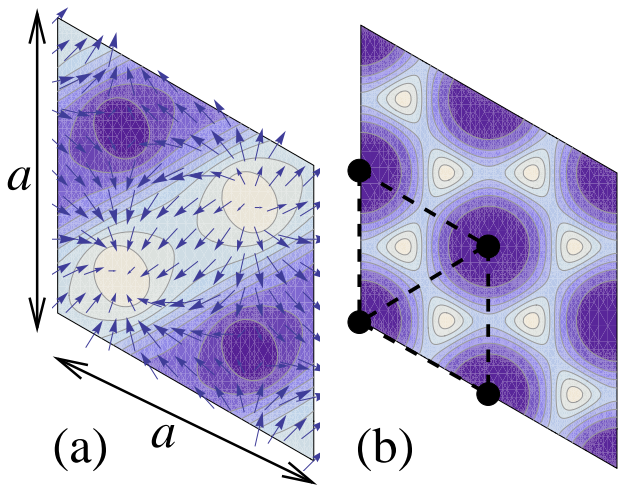}
\caption{ \label{fig:triangular} (a) Bloch vector and (b) flux density
  $n_\phi$ for the lower energy dressed state of the triangular
  optical flux lattice  (\ref{eq:vsq}). The local minima of the adiabatic
  energy are at the points where $n_\phi=0$, forming
a triangular lattice in the tight-binding limit (dark circles and dashed lines).}
\end{figure}
The properties of the lower energy dressed state are shown in
Fig.\ref{fig:triangular}.  The flux density is maximum on the sites of
a honeycomb lattice, and vanishes on the triangular lattice dual to
this. The tight-binding limit involves tunneling between the sites of
this dual triangular lattice, Fig.\ref{fig:square}(b).

The above optical potentials (\ref{eq:vsq},\ref{eq:vtriang}) can be
readily generalized to many other cases with non-zero mean flux.
(There are also many cases with zero mean, but non-zero local flux
density.) The central requirements to generate an optical flux lattice
are threefold. Firstly, the Raman laser ($M_{x,y}$) must generate
optical vortices. A 2D lattice of optical vortices can be formed from
a minimum of three travelling waves\cite{Masajada200121}.  The
resulting optical field is periodic, with an equal number of
(single-winding) vortices $N_{\rm v}$ and antivortices $N_{\rm av}$ in
a unit cell\cite{Note3}.
Secondly, the
species-dependent potential ($M_z$) must be non-zero at the cores of
these vortices, such that there is no degeneracy of the dressed states
at these points.  A small non-zero $M_z$ causes the cores of the
vortices to have the topology of ``merons''\cite{girvinleshouches}, in
which $\vec{n}({\bm r})$ sweeps over half of the Bloch sphere. For a
given meron, the sign of $M_z$ at its core times the sign of its
vorticity determines whether it contributes $+1/2$ or $-1/2$ a flux
quantum. The total number of flux quanta through the unit cell is
$N_\phi = N_{\rm v}^+ - N_{\rm av}^+$, where $N_{\rm v/av}^+$ is the
number of vortices/antivortices at which $M_z$ is positive. Thus, the
third requirement for a nonzero mean flux is that $M_z$ varies in
space such that $N_{\rm v}^+\neq N_{\rm av}^+$.

We have explored the properties of optical potentials generated by
simple laser patterns. An optical flux lattice can be generated using
just five travelling waves: three travelling waves of the Raman laser
($M_{x,y}$) to effect the vortex lattice, and a standing wave of the
species-dependent potential ($M_z$). [One such example is to remove
one of the four travelling waves from the Raman coupling in
(\ref{eq:vtriang}).] In all cases the local flux density is
inhomogeneous in space, in some even changing sign. Indeed, one can
show that, for smoothly varying optical fields, the flux density must
have at least $N_{\rm v} + N_{\rm av}$ zeroes in the unit
cell\cite{Note4}
The above cases (\ref{eq:vsq},\ref{eq:vtriang})
have non-negative flux density with the minimum number of zeroes.  For
three-, or more-, level systems, an optical flux lattice can have a
flux density that nowhere vanishes.  We have examples of optical
potentials that lead to such cases. However, these require more
involved laser configurations, so we do not pursue this here.

Having determined the properties of the optical flux lattices in the
adiabatic limit, we now turn to describe their bandstructures,
obtained from the eigenvalues of (\ref{eq:ham}).  The laser potentials
$\hat{M}_{\rm sq}$ (\ref{eq:vsq}) and $\hat{M}_{\rm tri}$
(\ref{eq:vtriang}) are clearly invariant under translations by the
respective lattice vectors ${\bm a}_{1,2}$.  In fact, they enjoy
higher translational symmetry, being invariant under the unitary
transformations
\begin{equation}
 \hat{T}_1 \equiv \hat{\sigma}_y e^{\frac{1}{2}{\bm a}_1\cdot
{\bm \nabla}}
\quad
 \hat{T}_2 \equiv \hat{\sigma}_x e^{\frac{1}{2}{\bm a}_2\cdot
{\bm \nabla}}
\end{equation}
which effect translations by $\frac{1}{2}{\bm a}_{1,2}$ and rotations
in spin-space. These operators do not commute, but satisfy
\begin{equation}
\hat{T}_2\hat{T}_1 = -\hat{T}_1\hat{T_2} \,.
\label{eq:magtrans}
\end{equation}
This indicates that they represent {\it magnetic} translations around
a region of space (enclosed by $\frac{1}{2} {\bm a}_1$ and
$\frac{1}{2} {\bm a}_2$) that contains $1/2$ a flux quantum.  As is
conventional in systems with magnetic translation
symmetry\cite{thoulesschern}, we define a magnetic unit cell that
encloses an integer number of flux: we choose ${\bm a}_1$, ${\bm
  a}_2/2$.  Writing the eigenvalues of the associated (commuting)
translation operators $\hat{T}_1^2$ and $\hat{T}_2$ as $e^{i{\bm
    k}\cdot {\bm a}_1}$ and $e^{i{\bm k}\cdot {\bm a}_2/2}$ defines
the Bloch wavevector ${\bm k}$ and the associated Brillouin zone.  The
additional symmetry $\hat{T}_1$ and the condition (\ref{eq:magtrans})
cause the energy spectrum $E_{\bm k}$ for all bands to be invariant
under ${\bm k}\cdot{\bm a}_2/2 \to {\bm k}\cdot{\bm a}_2/2 \pm \pi$.

For the square optical flux lattice (\ref{eq:vsq}), a solution of the
bandstructure shows that the lowest energy band does not overlap any
higher band for $V \gtrsim 0.1\hbar^2\kappa^2/m$.  The Chern
number\cite{thoulesschern} of this band is $1$, the sign being
reversed under an odd number of sign changes to the terms in
(\ref{eq:vsq}).  Thus, the lowest energy band is topologically
equivalent to the lowest Landau level of a charged particle in a
uniform magnetic field.
It is instructive to consider the bandstructure for $V \gg
\hbar^2\kappa^2/m$, when the variation in the adiabatic energy is
dominant, and the low energy bands are well described by a
tight-binding model\cite{jakschbh}. The minima of the adiabatic
potential form a square lattice, Fig.\ref{fig:square}(b).
Nearest-neighbour hopping on this square lattice leads to a model in
which each plaquette encloses $1/2$ a flux quantum.  The magnetic unit
cell contains two lattice sites, so there are two tight-binding
bands. The bands touch at two Dirac points\cite{moraisprl}, so one can
speak only of the Chern number of the two bands together. This total
Chern number is zero, consistent with the fact that this
nearest-neighbour tight-binding model is time-reversal symmetric\cite{Note5}.
In the physical model,
with $mV/\hbar^2\kappa^2$ large but finite, time-reversal symmetry is
broken by {\it next} nearest-neighbour hopping across diagonals of the
square lattice. This leads to closed loops around plaquettes which
contain $1/4$ of a flux quantum. This perturbation acts to split the
bands at the two Dirac points, and the two bands acquire Chern numbers
of $\pm 1$.

\begin{figure}
\includegraphics[width=0.9\columnwidth]{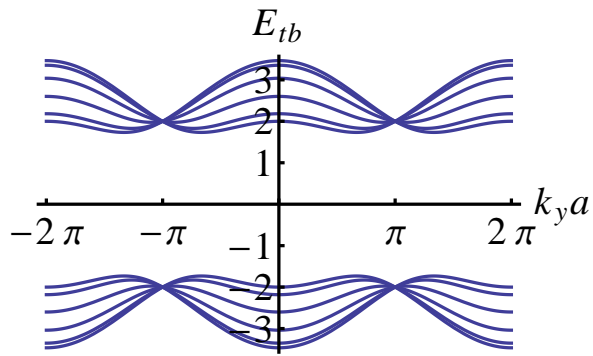}
\caption{ \label{fig:triband} Lowest energy bands for the triangular
  flux lattice (\protect\ref{eq:vtriang}) in the nearest-neighbour
  tight-binding limit (for uniform spacing of $-2\pi/\sqrt{3}\leq
  k_xa\leq 2\pi/\sqrt{3}$).  The energy $E_{tb}$ is relative to the
  atomic limit, in units of the nearest-neighbour hopping. The bands
  have Chern numbers $\pm 1$.}
\end{figure}

The bandstructure of the triangular optical flux lattice
(\ref{eq:vtriang}) has the same qualitative properties, the lowest
energy band having a Chern number of $1$.  In this case, time-reversal
symmetry is broken even in the tight binding limit with
nearest-neighbour hopping. The energy minima are at the sites of a
triangular lattice, Fig.\ref{fig:triangular}(b), the elementary
plaquettes of which enclose $1/4$ of a flux quantum.  The energy
spectrum of the resulting tight-binding model, shown for a convenient
gauge in Fig.\ref{fig:triband}, has two narrow bands that are well
separated in energy and have Chern numbers of $\pm 1$.

Optical flux lattices will allow experiments on ultracold gases to
explore many very interesting phenomena.  Since they lead to a lowest
energy band with non-zero Chern number, non-interacting fermions
filling this band (with one fermion per magnetic unit cell) will
exhibit the integer quantum Hall effect. Signatures of the resulting
chiral edge state could be observed in the density excitations
(collective modes), which will rotate with a handedness determined by
the sign of the Chern number.  The square lattice in the
nearest-neighbour tight binding limit also offers the possibility to
study fermionic Dirac physics.  Within mean-field theory, interacting
bosons loaded into the chiral band will develop vortex lattices with
very high flux density. These typically break translational symmetries
of the lattice\cite{mollercoopermft}.  Owing to the very high flux
density, it should be possible to reach a regime where the 2D boson
density is comparable to the mean flux density, $\bar{n}_\phi \sim
1/\lambda^2$, where strongly correlated fractional quantum Hall states
of bosons\cite{cwg,advances} or related states on
lattices\cite{mollercooper-cf} can appear.  A leading candidate is the
$\nu=1/2$ bosonic Laughlin state on the triangular lattice
(\ref{eq:vtriang}), for which the lowest energy chiral band is narrow
and well-separated from higher bands.  There is $1/2$ a flux quantum
per lattice site, so the Laughlin state appears at $1/4$ filling.  It
will be interesting also to explore strong correlation phenomena in 3D
settings, with an optical flux lattice providing net flux in one
direction.

\vskip0cm

\acknowledgments{I am grateful to Jean Dalibard for many helpful
  comments. This work was supported by EPSRC Grant EP/F032773/1.}


\end{document}